\documentclass[aps,prl,showpacs,twocolumn,showkeys,amssymb,amsmath,nobibnotes,superscriptaddress]{revtex4-1}
\usepackage[colorlinks=true,linkcolor=red]{hyperref}
\usepackage{amsmath}
\usepackage{amsfonts}
\usepackage{amssymb}
\usepackage{times}
\usepackage{color}
\usepackage{graphicx}
\usepackage[sort&compress]{natbib}

\newcommand{\sx}{\sigma^x}

\newcommand{\sz}{\sigma^z}

\newcommand{\taup}{{\tau^\prime}}

\newcommand{\tr}{\,\text{Tr}\,}

\begin{document}

\title{Collective excitations and marginal stability of quantum Ising spin glasses}
\author{A. Andreanov}
\affiliation{The Abdus Salam ICTP - 
Strada Costiera 11, 34151, Trieste, Italy}
\author{M. M\"uller}
\affiliation{The Abdus Salam ICTP - 
Strada Costiera 11, 34151, Trieste, Italy}

\begin{abstract}
We solve the Sherrington-Kirkpatrick (SK) model in a transverse field $\Gamma$ deep in its quantum glass phase at zero temperature. We show that the glass phase is critical everywhere, exhibiting collective excitations with a gapless Ohmic spectral function. Using an effective potential approach, we interpret the latter as being due to disordered spin waves which behave as weakly coupled, underdamped harmonic oscillators. In the limit of small transverse field $\Gamma$ the low frequency tail of the spectrum tends to a universal limit independent of $\Gamma$. 
\end{abstract}

\date{\today}

\maketitle

Spin glasses are canonical representatives of a wide class of complex disordered systems where competing (e.g. ferro- versus antiferromagnetic) interactions induce frustration, suppressing the emergence of simple ordering patterns. Nevertheless the interactions induce a phase transition from a disordered paramagnetic to a glassy "ordered" state at low temperatures. The emerging glass phase features several remarkable properties, which have been studied in great detail in classical glasses: despite of long range correlations there is no regular spatial order. Furthermore, the free energy landscape is very rough containing a large number of local minima, separated by high barriers~\cite{mezard1987spin}. The latter entails the  breaking of ergodicity and intriguing long time out-of-equilibrium phenomena. Classical glass states feature criticality~\cite{de1983eigenvalues,fisher1986ordered} in the sense that spin-spin correlations are power law correlated, despite the absence of a broken continuous symmetry. In systems with long range interactions (e.g., in Coulomb glasses, or in the SK model) this criticality is reflected by a pseudogap in the distribution of local fields at low temperatures~\cite{efros1975coulomb,pankov2005nonlinear,mueller2004glass}.

It is important to understand how such frustrated, critical systems behave in the presence of quantum fluctuations, and how the latter influence the dynamics and the relevant low energy excitations, both close to quantum glass transitions, as well as deep in quantum glass phases. Experimentally, these questions are directly relevant for quantum spin glasses such as the compound LiHo$_x$Y$_{1-x}$F$_4$, where quantum fluctuations can be tuned by an external magnetic field~\cite{wu1991classical,schechter2006quantum,schechter2008liho}, Coulomb-frustrated semiconductors close to a metal insulator transition~\cite{davies1982electron,davies1984properties,efros1975coulomb,pastor1999melting,mueller2007mean}, flux-frustrated Josephson junction arrays~\cite{teitel1983josephson,halsey1985josephson,vinokur1987set,gupta1998glassiness,choi1987glassy}, proton glasses (systems with frustrated ferroelectric interactions)~\cite{courtens1984short,pirc1985tunneling,feng2006quantum}, many-body cavity QED~\cite{gopalakrishnan2009emergent,gopalakrishnan2011frustration,strack2011dicke,strackmueller2012}, etc. Strong enough quantum fluctuations induce tunneling below the barriers and restore ergodicity, producing a quantum glass transition~\cite{bray1980replica,fedorov1986quantum}, which has been the subject of many investigations~\cite{klemm1979quantum,chakrabarti1981critical,pirc1985tunneling,banerjee1994model,feng2006quantum,schechter2007quantum}.

In the past many theoretical studies on quantum glasses have focused on short range interacting systems and the enhanced relevance of Griffith effects on the low frequency dynamics in short range spin glasses~\cite{young1998spin}. However, several of the above mentioned condensed matter realizations of frustrated systems feature longer ranged interactions and may be approached theoretically from the limit of mean field models, such as quantum rotors~\cite{ye1993solvable}, $SU(N)$ Heisenberg spin glass~\cite{georges2000mean,*georges2001quantum} and the transverse field Ising spin glass~\cite{fedorov1986quantum,pirc1985tunneling,rozenberg1998dynamics,goldschmidt1990ising,arrachea2001dynamical,thirumalai1989infinite,miller1993zero}. For the mean field version of the latter, the behavior at the quantum phase transition is well understood~\cite{miller1993zero}, however little is known about the quantum glass phase except in the vicinity of criticality. Based on a Landau expansion, the glass phase has been conjectured to be critical~\cite{sachdev1995quantum} with a spectral function behaving as $|\omega|$ at low frequencies. Exact diagonalization in small systems~\cite{arrachea2001dynamical} has indeed shown such a trend. The quantum glass phase is expected to have similar complexity as the classical spin glass with multi-valley free energy landscape~\cite{usadel1991phase,goldschmidt1990ising,rozenberg1998dynamics}, but a formal description has remained elusive. So far, tractable models could be solved exactly only in the limit of a large number of vector/rotor components, in which the complexity and criticality captured by full replica symmetry breaking is lost. The latter phenomena are however among the most intriguing aspects of quantum glasses with critical phases. In contrast, quantum systems with closer similarity to structural glasses often exhibit discontinuous quantum glass transitions and non-critical glass phases~\cite{biroli2001quantum}.

\begin{figure}[h]
\includegraphics[width=\columnwidth]{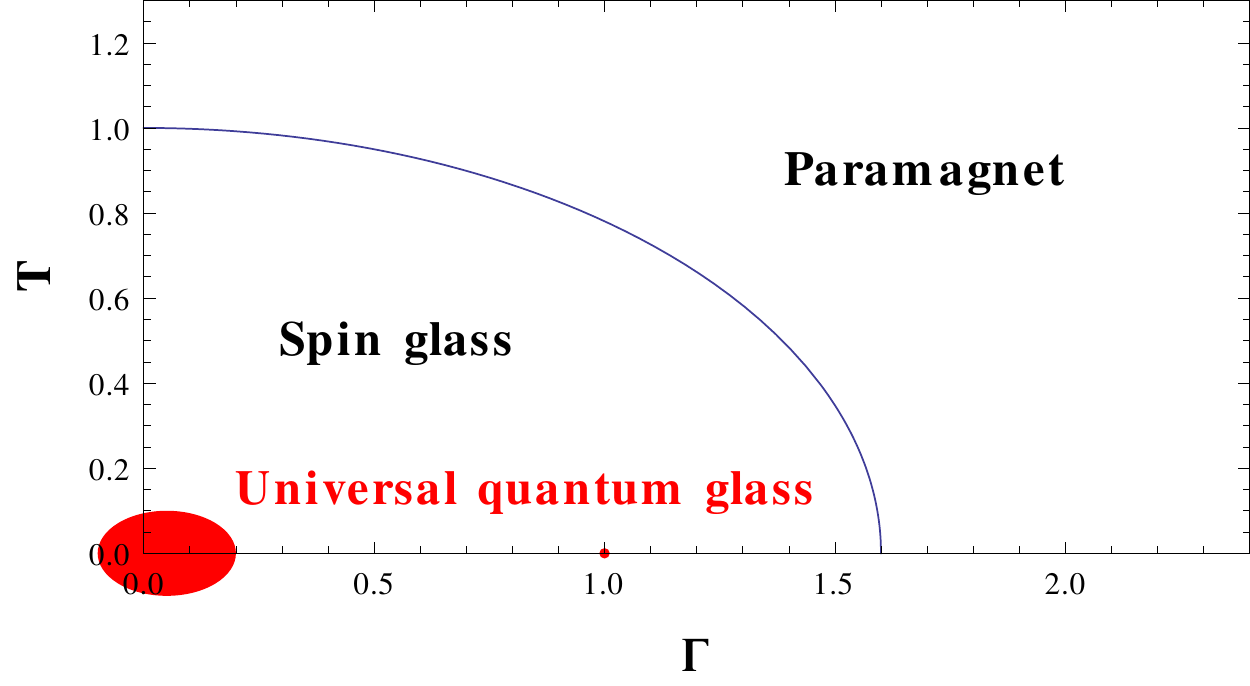}
\caption{Phase diagram of the mean field quantum spin glass. The deep quantum glass (red dot) exhibits gapless collective excitations with a low frequency spectrum {\em independent} of the transverse field $\Gamma$.}
\label{fig:phase_d}
\end{figure}

The present paper provides the missing link between the quantum phase transition and the deep quantum glass phase. We focus on Ising spin glasses, which combine a continuous glass transition with a critical glass phase and non-trivial ergodicity breaking. This model acts as a prototype for many glasses with long range interacting, discrete degrees of freedom, such as localized electrons, electric dipoles etc. Moreover, it has been recently been pointed out that the mean field model may be realized faithfully in random laser cavities, where a multitude of modes provide random long range couplings between trapped atoms, as in the Dicke model~\cite{strack2011dicke}. We study the quantum glass phase of  the Sherrington-Kirkpatrick model in a transverse field~\cite{chakrabarti1981critical}, i.e., the mean field version of the Ising quantum spin glass,
\begin{equation}
\label{model_ham}
\mathcal{H}=-\sum\limits_{i<j}J_{ij}\sz_i\sz_j-\Gamma\sum\limits_i\sx_i
\end{equation}
where $\sx$ and $\sz$ are Pauli operators. Every spin interacts with all the others, and the couplings $J_{ij}$ are random Gaussian variables of zero mean and variance $J^2/N$. The quantum fluctuations are tuned by the transverse field $\Gamma$. The phase diagram and the deep quantum glass regime of particular interest to us are shown in Fig.~\ref{fig:phase_d}. In the classical limit $\Gamma=0$ a glass transition takes place at $T_c=J$. As has long been known, it is connected by a line of continuous transitions to the quantum glass transition at $\Gamma_c(T=0)\approx 1.52 J$~\cite{read1995landau,miller1993zero}. Below we use units with $J=1$, and restore $J$ occasionally for clarity. 

\emph{Replica solution} - We first solve the model by the replica approach, and then interpret its features with the physically more transparent effective potential (TAP) method~\cite{thouless1977solution,biroli2001quantum}. The disorder average of the free energy is carried out using the replica trick following Ref.~\cite{bray1980replica}, reducing the problem to an effective, self-consistent single spin model: 
\begin{widetext}
\begin{gather}
\beta F = \frac{(\beta J)^2}{2}\sum\limits_{b\neq 1}Q_{1b}^2 + \frac{1}{4}\iint_0^\beta d\tau d\taup Q_{aa}^2(\tau,\taup) - \lim_{n\to 0} \left[ \frac{1}{n}\log\tr\,\text{T} e^{\mathcal{S}_\text{eff}}\right],\notag\\
\mathcal{S}_\text{eff} = \iint\limits_0^\beta d\tau d\taup\left[\sum\limits_{a<b}Q_{ab}\sz_a(\tau)\sz_b(\taup) + \frac{1}{2}\sum\limits_a Q_{aa}(\tau,\taup)\sz_a(\tau)\sz_a(\taup)\right] + \Gamma\sum\limits_a\int\limits_0^\beta d\tau\,\sx_a(\tau),\\
Q_{ab} = \langle\sz_a\sz_b\rangle\to q(x);\qquad Q_{aa}(\tau,\taup)=\langle\text{T}\,\sz_a(\tau)\sz_a(\taup)\rangle = C_{\tau-\taup} = R_{\tau-\taup}+q(1).\notag
\end{gather}
\end{widetext}
The saddle point values of the off-diagonal $Q_{ab}$ are time independent~\cite{cugliandolo2001imaginary} and have an ultrametric structure parameterized by the monotonous function~\cite{parisi1980sequence,*parisi1980order,goldschmidt1990ising,usadel1991phase} $0\leq q(x)\leq 1$, with $x\in[0,1]$ measuring the distance between replica in phase space, and thus being a proxy of a time scale in the aging regime. $R_{\tau-\taup}$ is the connected part of the replica-diagonal spin correlator $Q_{aa}(\tau,\taup)$, which tends to the Edwards-Anderson parameter $q_\textrm{EA}\equiv q(1)$ at large time separations at $T=0$. Assuming a continuous function $q(x)$, the self-consistency problem is equivalent to the solution of the diffusion-like equations~\cite{sommers1984distribution,thomsen1986local}:
\begin{gather}
\dot{m}(y,x) = -\frac{\dot{q}(x)}{2}\left[m^{\prime\prime}(y,x)+2\beta x\, m(y,x)\,m^\prime(y,x)\right],\notag\\
\dot{P}(y,x) = -\frac{\dot{q}(x)}{2}\left[P^{\prime\prime}(y,x)-2\beta x\, (m(y,x)\,P(y,x))^\prime\right],\notag\\
q(x) = \int dy\,P(y,x)\,m^2(y,x), \label{eqn:ppde}
\end{gather}
where dots and primes denote derivatives with respect to $x$ and $y$, respectively.  $P(y,x)$ is the distribution of frozen exchange fields $y$, 
averaged over time scales corresponding to phase space distances $x$, with $P(y,0)=\delta(y)$. Likewise, 
$m(y,x)$ is the magnetization of a spin in the presence of a frozen field $y$ on that time scale. Short-time observables are described by $x=1$, of which the local field distribution $P(y, 1)$ and the spin-spin correlator $R_{\tau}$ will be of particular interest. The difference to the classical problem lies in the modified set of boundary conditions which read: $m(y,x=1)=\langle\sz\rangle_{\mathcal{S}(y)}$ with the local action 
\begin{gather}
\label{eqn:ppde-ic}
\mathcal{S}(y) = \frac{1}{2}\iint_{0}^\beta d\tau d\tau' \,\sz_{\tau} R_{\tau-\tau'}\sz_{\tau'} + \int_0^\beta d\tau (y\,\sz_{\tau} + \Gamma\,\sx_{\tau}).
\end{gather}
The system of equations is closed by the self-consistency relation 
\begin{gather}
\label{eqn:sc_R}
R_{\tau-\tau'}= \int_y P(y,1)\langle\text{T}\,\sz_{\tau}\sz_{\tau'}\rangle_{\mathcal{S}(y)}-q(1).
\end{gather}

\emph{Solution at $T=0$} - We now focus on the $T=0$ behavior. Like at any quantum critical point, the gap closes at the transition~\cite{miller1993zero}. However, it was found within replica symmetric Landau theory~\cite{read1995landau} that the gap remains closed in its vicinity, with an Ohmic spectrum at small frequencies. Similar behavior was found in the analysis of a rotor model, where in the limit of $M\to\infty$ components~\cite{ye1993solvable} the replica symmetry is not broken.
Here we show that in the Ising case (believed to be the $M=1$ limit of the rotor model) full replica symmetry occurs, and that the latter guarantees that the gaplessness extends into the whole glass phase. 
We point out that this phenomenology contrasts with that of the exactly solvable model SU($N\to\infty$) Heisenberg spin glass~\cite{read1995landau,georges2000mean,georges2001quantum}, which exhibits a random first order transition with distinct dynamic freezing and thermodynamic glass transitions. Its  thermodynamically dominant states are gaped and are thus  very different from the states obtained in the Ising limit $N\to 1$ analyzed here.

The spectral function is encoded in the 
Fourier transform $R_\omega$ of the average spin correlator $R_\tau$, which we analyze following Miller and Huse~\cite{miller1993zero}. Representing the action $\mathcal{S}(y)$ with fermions, we can expand the spin-spin correlator into a power series in $R_\omega$
\begin{equation}
\label{eqn:susceptibility}
\langle\sz_{-\omega}\sz_\omega\rangle_{\mathcal{S}(y)} \equiv \chi_\omega(y) = \dfrac{\Pi_\omega(y)}{1 - R_\omega\Pi_\omega(y)},
\end{equation}
where $\Pi_\omega(y)$ is the proper polarizability~\cite{fetter2003quantum}, itself a functional of $R_\omega$. However, $\Pi$ remains analytic at small $\omega$, $\Pi_\omega\sim\Pi_0-a \omega^2$, even when $R_\omega$ turns non-analytic in the glass phase. The latter is a consequence of the marginal stability,
\begin{gather}
\label{eqn:ms}
\int_0^\beta dy P(y,1)(m^\prime)^2(y,1) = 1,
\end{gather}
which is implied by Eq.~\eqref{eqn:ppde}. Indeed, with the small frequency expansion $R_\omega=R_0+\delta R_\omega$, and noting that $m'(y,1) = \chi_{\omega\to 0}(y)$ as well as  $\chi_\omega(y)  = \chi_0(y) +\chi_0^2(y) \delta R_\omega+O(\omega^2)$, Eqs.~(\ref{eqn:sc_R}-\ref{eqn:ms}) require that $\delta R_\omega^2 \sim \omega^2$. This implies the non-analyticity $R_{\omega\to0}=R_0 - B|\omega|$ in the low frequency correlator. Upon analytic continuation, the spin spectral function 
\begin{eqnarray}
A(\omega\to0)\equiv \frac{1}{\pi N}\sum_i{\rm Im} \langle s^z_i(\omega)s_i^z(-\omega)\rangle\vert_{\omega\to\omega - i\delta} = 
\frac{B\omega}{\pi},\quad
\label{eqn:spectralfct}
\end{eqnarray}  
is found to be \emph{always} gapless and Ohmic, even deep in the Ising glass phase. Remarkably, as we will derive below, $B$ becomes $\Gamma$-independent as $\Gamma\to 0$. The marginal stability and the related gaplessness of the replicon mode are both natural by-products of continuous replica symmetry breaking in mean field glasses. They replace the role of Goldstone modes in systems with a broken continuous symmetry. Physically, the gapless excitations (\ref{eqn:spectralfct}) arise due to the presence of many metastable states which are close to the ground state and connected to it via nearly flat directions in the energy landscape. The transverse field hybridizes the nearby classical states, which gives rise to very low-lying collective spin excitations.

\emph{Deep glass phase} - To obtain quantitative results we have solved the full self-consistency problem, focusing on the deep quantum glass phase where $J\gg \Gamma\gg T\to 0$. The limit $T\to 0$ is taken by replacing the variable $x\in [0,1]$ by $\beta_{\rm eff}\equiv \beta x \in [0,\infty]$, which has the interpretation of an inverse effective temperature in the aging dynamics~\cite{cugliandolo1993analytical,*cugliandolo1994out}. In the limit $\Gamma\ll J$, the flow of Eqs.~(\ref{eqn:ppde}) is attracted to a scaling regime where $dq/d\beta_{\rm eff} = c(\beta_{\rm eff})/\beta^{3}_{\rm eff}$,  $m(x,y)\to \tilde m(\beta_{\rm eff} y)$, and $P(x,y)\to \beta_{\rm eff} \tilde p(\beta_{\rm eff} y)$, which holds for $1/J\ll\beta_{\rm eff}\ll 1/\Gamma$ and $y\ll J$.  Here, $c(\beta_{\rm eff})\to 0.411$, and $\tilde m$ and $\tilde p$ are the same fixed point functions that appear in the classical low $T$ limit~\cite{pankov2006low}. In particular,
\begin{gather}
\tilde{p}(\beta_{\rm eff} y)\approx\left\{
\begin{matrix}
\alpha|\beta_{\rm eff} y| & 1\ll |\beta_{\rm eff} y| ,\\
{\rm const.} & |\beta_{\rm eff} y|\lesssim 1,
\end{matrix}
\right.
\end{gather}
which displays a linear pseudogap with slope $\alpha=0.301$ in the distribution of frozen fields, smeared on the scale $y\sim 1/\beta_{\rm eff}$. 

\begin{figure}
\includegraphics[width=\columnwidth]{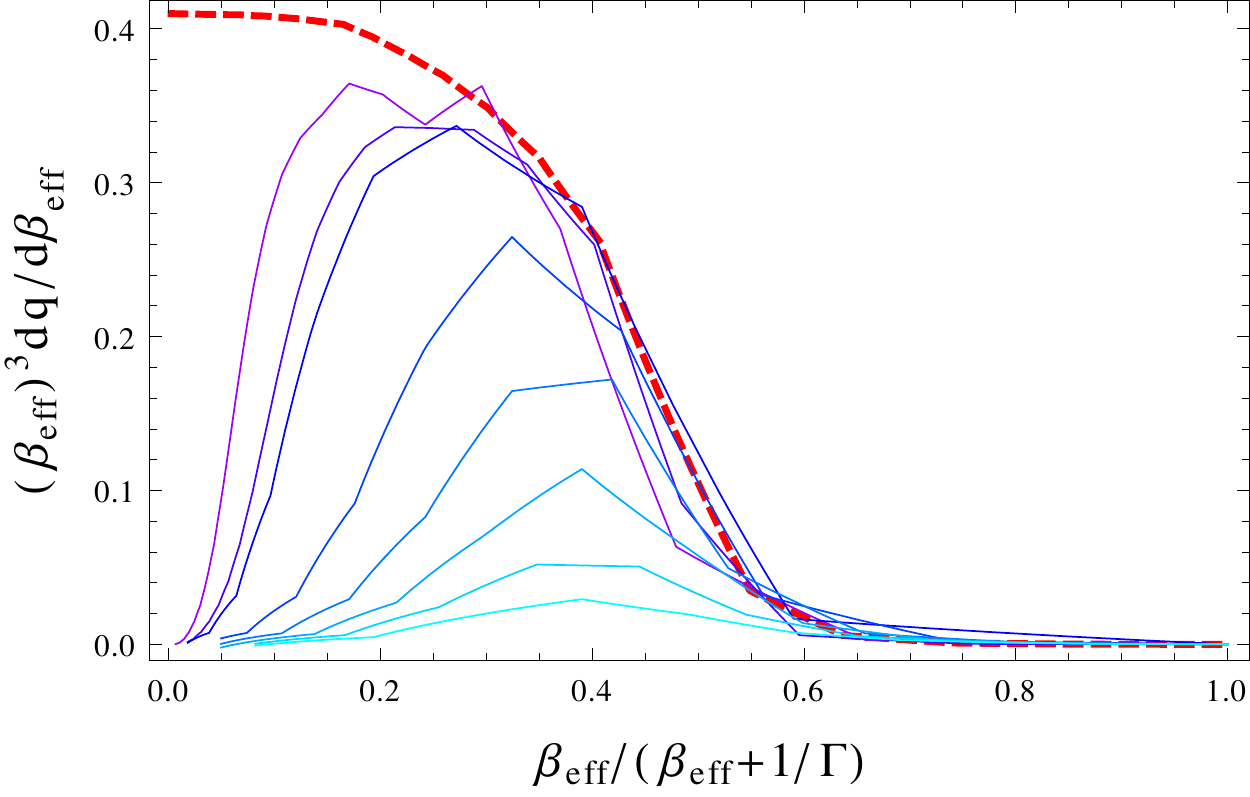}
\caption{Rescaled derivative of the overlap function, $c=\beta_{\rm eff}^3 dq/d\beta_{\rm eff}$ as a function of $u\equiv \beta_{\rm eff}/(\beta_{\rm eff}+1/\Gamma)$, extracted from the full solution  for finite $\Gamma$ (solid curves). The dashed curve is the solution of the asymptotically exact scaling anzats for $\Gamma/\Gamma_c\to 0$ ~\cite{am_full}. The transverse field $\Gamma$ takes values (bottom to top) $\Gamma/\Gamma_c= 0.05,0.08,0.1,0.2,0.3,0.4,0.5,0.6$. Convergence to the solution is slower~\cite{am_full,pankov2006low} and numerical instabilities are more pronounced for lower values of $\Gamma$, $\Gamma\lesssim0.1\Gamma_c$ producing rougher curves.  The scaling regime obtains at $\beta_{\rm eff}\ll 1/\Gamma$ ($c\to 0.411$). It reflects the ultrametric structure of phase space and an ensuing self-similar structure of dynamics deep in the aging regime~\cite{pankov2006low, mueller2007mean}. $q(x)$ reaches its plateau value $q_{\rm EA}$ at $\beta_{\rm eff}^c= x_c/T\approx  0.5/\Gamma$ where $c\to 0$.}
\label{fig:collapse}
\end{figure}

\begin{figure}
\includegraphics[width=\columnwidth]{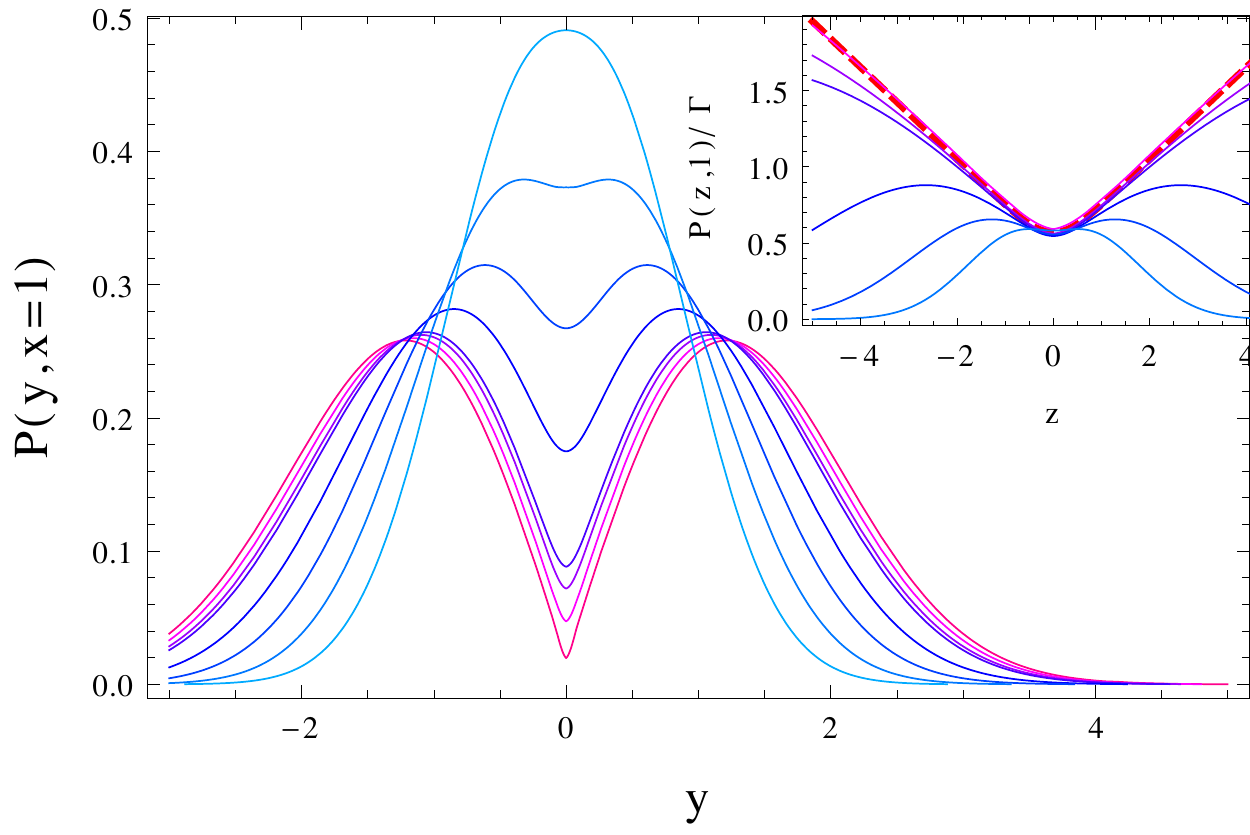}
\caption{Opening of the linear pseudogap in the distribution of frozen fields in the glassy ground state at $T=0$, $P(y,1)$, deeper and deeper into the quantum glass ($\Gamma/\Gamma_c\to 0$ from top to bottom, same values as in Fig.~\ref{fig:collapse}). The inset shows the rescaled distribution $P(y=z \Gamma)/\Gamma$ (solid lines), compared to the asymptotically exact scaling function $\tilde{p}(z)$ (dashed line)~\cite{am_full}. }
\label{fig:p_gap}
\end{figure}

For $\beta_{\rm eff}\gtrsim 1/\Gamma$, the overlap 
ceases to scale. The derivative drops [$c(\beta_{\rm eff})\to 0$, cf. Fig.~\ref{fig:collapse}], $q(\beta_{\rm eff})$ reaches the constant value $q_{\rm EA}$, and the flow of $P$ and $m$ freeze to their short time forms. This yields the physical result that the frozen field distribution in typical glass states, $P(y,1)$ has a pseudogap, which is smeared on the scale $\Gamma$, as one may expect from stability arguments. The evolution of the pseudogap upon entering deeper into the quantum glass is shown in Fig.~\ref{fig:p_gap}.

In the limit $\Gamma\ll J$, $\Gamma$ is the only relevant dynamic energy scale, while $J$ merely determines the width of the distribution of frozen fields. We demonstrated this by analyzing the natural scaling anzats: {\em (i)} for $\beta_{\rm eff}\gg 1/J$, $c(\beta_{\rm eff}) = \hat c(u)$ is only a function of $u=\beta_{\rm eff}/(\beta_{\rm eff}+1/\Gamma)$; {\em (ii)}   $R(\omega)= \Gamma \hat r(\omega/\Gamma)$ and {\em (iii)} $P(y,1)=\Gamma \hat p(y/\Gamma)$ [and $P=(\Gamma/u) \hat p(y u/\Gamma,u)$, $m= \hat m(y u/\Gamma,u)$]. One then verifies that  $\hat c, \hat r, \hat p$ and $\hat m$ satisfy self-consistent equations which are {\em independent} of $\Gamma\ll \Gamma_c$. They admit a solution to which the full solutions of Eqs.~\eqref{eqn:ppde}-\eqref{eqn:sc_R} converge as $\Gamma\to 0$, as shown in Figs.~\ref{fig:collapse} and \ref{fig:p_gap}.
This solution implies a non-trivial scaling of the dynamic properties of the glass; in particular it entails the remarkable result that the coefficient $B$ of the Ohmic spectral function~\eqref{eqn:spectralfct} tends to a constant $B\approx 0.59/J^2$ as $\Gamma\to 0$.     
 
\emph{Physical interpretation} - We now interpret the above results physically, based on the effective potential (TAP) approach~\cite{thouless1977solution}.
The $\Gamma$-independence of the low frequency tail of the spectral function, predicted by the replica solution for the deep glass phase, will serve us as a benchmark. Using the method of Refs.~\cite{plefka1982convergence,biroli2001quantum} we construct the Gibbs potential $G[\mathbf{m},C_\tau]$ describing the free energy of the system constrained to have a magnetization pattern $\{m_i\}$ and global autocorrelation function $C_\tau=1/N\sum_i\langle\sz_i(\tau)\sz_i(0)\rangle$ (at $T=0$):
\begin{eqnarray}
G[\mathbf{m},C_\tau] =& \sum_i G_0[m_i,C_\tau] - \sum\limits_{i<j}J_{ij}\,m_i\,m_j \nonumber\\
& - \frac{1}{4}NJ^2\int\limits_0^\infty d\tau\left(C_\tau - q_\text{EA}\right)^2.
\label{eqn:tap_G}
\end{eqnarray}
Here $q_\textrm{EA}=1/N\sum_im_i^2$ and $G_0$ is the free energy of single, constrained spin. The magnetizations of local minima $m_i=\langle\sz_i\rangle$ are computed self-consistently via $\delta G/\delta m_i = 0$:
\begin{equation}
\left.\dfrac{\partial G_0}{\partial m_i}\right\vert_C - \sum_j J_{ij}m_j + J^2\,m_i\,\chi^0 = 0,
\label{eqn:tap_m}
\end{equation}
where $\chi^0=\int_0^\infty d\tau(C_\tau - q_\textrm{EA})= 1/N\sum_i\chi_i$ is the static susceptibility. However, for quantum problems Eq.~\eqref{eqn:tap_m} is not closed since $G_0$ depends on the global autocorrelation function $C(\tau)$, which has to be evaluated self-consistently, too~\cite{biroli2001quantum}. Since this exact formalism is too involved to yield direct physical insight, we approximate the static susceptibilities $\chi_i$ and the local functional $G_0$ by those of single spins, whose magnetization $m_i$ is constrained by an auxiliary \emph{static} field:
\begin{gather}
G_0[m_i] = -\Gamma(1-m_i^2)^{1/2},\quad 
\overline{\chi}_i 
= \dfrac{(1-m_i^2)^{3/2}}{\Gamma}.
\end{gather}
This approximation is similar but not identical to the "static approximation" employed in replica approaches to quantum spin glasses~\cite{thirumalai1989infinite}. It overestimates the susceptibility to longitudinal fields, enhancing the stability of the glass. However, it reproduces qualitatively the results of the rigorous replica theory, furnishing a complementary physical picture.  

Collective excitations in a local minimum are governed by the curvature of the energy landscape, i.e. by the Hessian 
\begin{equation*}
\mathcal{H}_{ij} = \dfrac{\delta^2 G}{\delta m_i\delta m_j} = -J_{ij} + \left[\frac{1}{\overline{\chi}_i} + J^2\langle\overline{\chi}\rangle\right]\delta_{ij},
\end{equation*}
where $\langle\overline{\chi}^k\rangle=1/N\sum_i\overline{\chi}_i^k.$ The replica theory assures that the glass phase is marginal. Here this translates into a gapless spectrum of eigenvalues of $\mathcal{H}_{ij}$, which requires~\cite{bray1979evidence}
\begin{gather}
\label{eqn:ms2}
J^2 \langle\overline{\chi}^2\rangle = 1.
\end{gather}
This is the natural analog of Eq.~\eqref{eqn:ms}. Under this condition, the density of eigenvalues of the Hessian $\lambda$ is given by 
\begin{equation*}
\rho(\lambda) = \dfrac{\sqrt{\lambda}}{\pi J^3\sqrt{\langle\overline{\chi}^3\rangle}}\sim\dfrac{\sqrt{\lambda\Gamma}}{J^2},\quad\text{for}\,\,\lambda_*\sim\Gamma\ll J.
\end{equation*}

To establish the link with the spin spectral density (\ref{eqn:spectralfct}) we interpret the low energy normal modes of $\mathcal{H}$ as weakly interacting~\footnote{The weakness of interaction between the modes is supported by parametric smallness of inelastic scattering rate for $\omega<\Gamma$~\cite{am_full}.} harmonic oscillators with spring constant $\lambda$, an effective mass scaling as $M\sim1/\Gamma$, and thus an eigenfrequency $\omega(\lambda)=\sqrt{\lambda/M}$. Hence, the density of modes is 
\begin{equation*}
\rho(\omega)=\int d\lambda\,\rho(\lambda)\,\delta(\omega-\omega(\lambda))\sim\dfrac{\omega^2}{\Gamma J^2},\quad\text{for}\,\,\Gamma\ll J.
\end{equation*}
Using the mean square displacement $\langle x^2\rangle_\omega=1/M \omega \sim\Gamma/\omega,$ the spectral function results as, $A(\omega)\approx\rho(\omega)\langle x^2\rangle_\omega\sim \omega/J^2$ for $\omega\lesssim\Gamma$. Thus, these qualitative arguments are seen to reproduce correctly the Ohmic spectrum, its frequency range, and the $\Gamma$-independent coefficient of the replica solution (\ref{eqn:spectralfct}). The latter is non-trivial given that both the mode density and the kinetic energy of the soft modes do depend on $\Gamma$. This non-trivial check makes us confident that the physical picture of a set of gapless, underdamped collective harmonic oscillators is the correct interpretation of the low energy spin excitations in this quantum glass. It is interesting to note that an analogous reasoning for spin glasses with metallic background leads to a similar picture, however with overdamped oscillators, and a spectral function growing as $A(\omega)\sim |\omega|^{1/2}$, again in agreement with replica theory~\cite{sachdev1995quantum,am_full}.

The appeal and potential of the TAP approach lies not only in the physical picture it provides for the collective excitations, but also in the fact that it lends itself to generalizations to finite dimensional glasses with long, but not infinite range interactions~\cite{mueller2007collective}. While in such more realistic models criticality might not exist at all length scales, collective modes are expected to persist down to very small energy scales, if the interaction range is large. Those may play an important role in activated transport of localized charge glasses~\cite{mueller2007collective}, or induce non-Fermi liquid corrections in metallic glasses~\cite{sengupta1995non,dalidovich2002landau}. It would be interesting to understand the spatial nature of these excitations, and to study whether and how the very low frequency spectrum eventually becomes dominated by more localized, droplet excitations, as they occur in quantum Griffith phases. In the context of electronic glasses, many further interesting questions arise, in particular as to the interplay of glassy ordering with quantum phenomena  such as (disordered) superfluidity and Bose glasses~\cite{yu2012mean}, as well as Anderson localization~\cite{strackmueller2012}.  

We would like to thank D. Carpentier, L. Cugliandolo, S. Florens and P. Strack for many useful discussions. We are grateful to P. Strack for a careful reading of the manuscript.

\bibliography{tfsk}{}

\end{document}